%% file: 2022_wifs.tex
\documentclass[conference]{IEEEtran}
\usepackage[utf8]{inputenc}
\usepackage{environ}
\usepackage{amsmath}
\usepackage{amssymb}
\usepackage{graphicx}
\usepackage{placeins}
\usepackage[table,xcdraw,usenames,dvipsnames]{xcolor}
\usepackage{dsfont}

\usepackage{hyperref}

\definecolor{source}{HTML}{154360}
\definecolor{target}{HTML}{17A589}

\usepackage{makecell}

\newcommand\numberthis{\addtocounter{equation}{1}\tag{\theequation}}

\NewEnviron{myequation*}{%
    \begin{equation*}
    \scalebox{1.2}{$\BODY$}
    \end{equation*}
    }

\NewEnviron{myequation}{%
\begin{equation}
\scalebox{1.2}{$\BODY$}
\end{equation}
}

\NewEnviron{myalign}{%
\begin{align*}
\scalebox{1.2}{\BODY}
\end{align*}
}

\ifCLASSINFOpdf
\else
\fi
\usepackage{algorithm, algpseudocode}
\algrenewcommand\algorithmicrequire{\textbf{Input:}}
\algrenewcommand\algorithmicensure{\textbf{Output:}}

\usepackage{stfloats}
\begin{document}
%
\title{Using Set Covering to Generate Databases for Holistic Steganalysis }

\author{\IEEEauthorblockN{Rony Abecidan}
\IEEEauthorblockA{Univ. Lille, CNRS, Centrale Lille, \\ UMR 9189 CRIStAL,\\ F-59000 Lille, France\\
Email: rony.abecidan@univ-lille.fr}

\and

\IEEEauthorblockN{Vincent Itier}
\IEEEauthorblockA{IMT Nord Europe, Institut Mines-Télécom,\\ Centre for Digital Systems, Univ. Lille, CNRS, Centrale Lille,\\ UMR 9189 CRIStAL,F-59000 Lille, France\\
Email: vincent.itier@imt-nord-europe.fr}

\linebreakand
\and
\IEEEauthorblockN{Jérémie Boulanger}
\IEEEauthorblockA{Univ. Lille, CNRS, Centrale Lille, \\ UMR 9189 CRIStAL,\\ F-59000 Lille, France\\
Email:  jeremie.boulanger@univ-lille.fr}

\and
\IEEEauthorblockN{Patrick Bas}
\IEEEauthorblockA{Univ. Lille, CNRS, Centrale Lille, \\ UMR 9189 CRIStAL,\\ F-59000 Lille, France\\
Email:  patrick.bas@cnrs.fr}
\and
\IEEEauthorblockN{Tomáš Pevný}
\IEEEauthorblockA{Department of Computers and Engineering\\ Czech Technical University\\ Prague, Czech Republic\\
Email: pevnak@protonmail.ch}
}

%



\makeatletter
\newcommand{\linebreakand}{%
  \end{@IEEEauthorhalign}
  \hfill\mbox{}\par
  \mbox{}\hfill\begin{@IEEEauthorhalign}
}
\makeatother

\maketitle

\begin{figure}[b]
  \vspace{-0.3cm}
  \parbox{\hsize}{\em
  WIFS`2022, December, 12-16, 2022, Shanghai, China.
  XXX-X-XXXX-XXXX-X/XX/\$XX.00 \ \copyright 2021 IEEE.
  }\end{figure}

\newcommand\NPB[1]{{\color{green!70!black}{ \small { [PB: ~#1~]}}}}

\newcommand\NRA[1]{{\color{red!70!black}{ \small { [RA: ~#1~]}}}}

\newcommand\NVI[1]{{\color{blue!70!black}{ \small { [VI: ~#1~]}}}}

\newcommand\NJB[1]{{\color{magenta!70!black}{ \small { [JB: ~#1~]}}}}

\newcommand\QF[1]{\textsc{qf}#1}

\begin{abstract}
  Within an operational framework, covers used by a steganographer are likely to come from 
  different sensors and different processing pipelines than the ones used by researchers for training their steganalysis models. 
  Thus, a performance gap is unavoidable when it comes to out-of-distributions covers, an extremely frequent scenario called Cover Source Mismatch (CSM).
  Here, we explore a grid of processing pipelines to study the origins of 
  CSM, to better understand it, and to better tackle it.
  A set-covering greedy algorithm is used to select representative pipelines minimizing the maximum regret between the representative and the pipelines within the set. 
  Our main contribution is a methodology for generating relevant bases able to tackle operational
  CSM. Experimental validation highlights that, for a given number of training samples, 
  our set covering selection is a better strategy than
  selecting random pipelines or using all the available pipelines.
  Our analysis also shows that parameters as
  denoising, sharpening, and downsampling are very important to foster 
  diversity. Finally, different benchmarks for classical 
  and wild databases show the good generalization 
  property of the extracted databases. Additional resources are available at \href{https://github.com/RonyAbecidan/HolisticSteganalysisWithSetCovering}{github.com/RonyAbecidan/HolisticSteganalysisWithSetCovering}.

\end{abstract}


%
\IEEEpeerreviewmaketitle

\section{Introduction}
\label{sec:intro}

\input{introduction.tex}

\section{Source screening with set-covering}
\label{sec:formalization}
\input{formalization.tex}

\section{Practical Selection of Source Representatives}
\label{sec:experiments}
\input{experiments.tex}

\section{Benchmark on different databases}
\label{sec:tackling_csm}
\input{tackling_csm.tex}

\section{Conclusions and Perspectives}
\label{sec:conclusion}
\input{conclusion.tex}


\section*{Acknowledgments}

Our experiments were possible thanks to computing means of IDRIS through the resource allocation 2021-
AD011013285 assigned by GENCI. This work received funding from the European Union’s
Horizon 2020 research and innovation program under grant agreement No 101021687 (project “UNCOVER”) 
and the French Defense \& Innovation Agency. The work of Tomas Pevny was supported by Czech Ministry of Education 19-29680L.




\bibliographystyle{IEEEtran}
\bibliography{biblio}
%

%
%
%
%
%

\end{document}

%% file: introduction.tex




Cover-source mismatch (a.k.a. CSM) is well-known in modern steganalysis 
\cite{giboulot} \cite{csm_chaumont} \cite{csm_fridrich}.
In the literature, steganalysis models are commonly trained on controlled cover distributions
coming 
from BOSSBASE~\cite{bossbase} or ALASKABASE~\cite{alaskabase} for instance. Meanwhile, in operational steganalysis, it is rarely possible to guess the distributions to which the images belong. Cover distributions, also called cover sources, present a lot of diversity because of several factors such as 
the image acquisition device (camera, mobile phones, scanner, \emph{etc.}), the quality of the sensor, the settings of this device
(ISO, zoom, aperture, shutter time, etc.), the content captured (inside, outside, luminosity, 
level of details, \emph{etc.}), the post-processing step
(sharpening, denoising, white balance, gamma correction, cropping, \emph{etc.}),
and also the usual compression step ($8$-bit conversion, JPEG compression, \emph{etc.}).
These development pipelines correspond to a set of transformations associated with parameters impacting the statistics of the developed image before a potential embedding.
In this context, a cover source can be seen as a mix of two distributions: the content and
the noise. 
Previous studies have shown that the mismatch between two cover sources
is generally much more
fostered by the noise distribution than by that of the content~\cite{giboulot,quentin:hal-03694662}. 

More generally, it is shown in \cite{giboulot} that the
processing pipeline 
is the main perpetrator of CSM. Table \ref{fig:big_csm}   
shows how far two sources can mismatch if 
they differ only by one parameter value in their processing pipelines.
This set of transformations is commonly used for aesthetic and
compression purposes. In the steganalysis case, it is impacting significantly the noise distribution 
while keeping the semantics pristine. Even if machine learning schemes are effective for steganalysis
tasks, they are very often extremely sensitive to the very nature of the analyzed signal. This is why CSM might occur.

\begin{table}[b]
  \centering
  \begin{tabular}{|c|c|c|}
  \hline
  Train / Eval           & \textbf{No Denoising} & \textbf{Max Denoising} \\ \hline
  \textbf{No Denoising}  & 5                     & 77                     \\ \hline
  \textbf{Max Denoising} & 48                    & 0.18                   \\ \hline
  \end{tabular}
  \vspace*{0.3cm}
  \caption{$P_E$ matrix (Probability of Error in\%) between two mismatching sources only differing by a denoising factor in their processing
  pipelines.
  Messages are embedded with UERD\cite{uerd} with a payload of 1.5bpp. Training with a linear classifier on DCTr features \cite{dctr}. 
  Pipelines 5 and 167 in our directory.}
  \label{fig:big_csm}
\end{table}


To address this issue, several approaches have been designed. On one hand, atomistic approaches assume that the distribution of the test covers can be reproduced.
It requires to have access to covers close to the test ones in terms of distribution.
Then, it is possible to create a batch of classifiers trained on specific sources 
and use them accordingly. For example, in \cite{csm_fridrich}, the authors
propose 
to pick up the classifier trained on the closest distribution to the test one.
Whereas, in \cite{csm_chaumont}, the authors use an ensemble 
classifier to face the CSM problem. 

On the other hand, there are holistic approaches for which a mixture of relevant cover distributions 
should help to cope with the heterogeneity of the test set~\cite{csm_fridrich}. The purpose of these methods is to bring a lot of
information and diversity to the dataset. For example, to be the most 
representative possible in terms of content and noise,
 pictures from ALASKABASE~ \cite{alaskabase} are coming
from 479 different cameras with various ISO ranging from 16 to 51200.

Generally, the holistic approach is more suitable because it does not require too much assumption about the cover distribution.
In this work, we put ourselves in a realistic scenario where we do not know anything about the distribution
of the test covers. Hence, we propose a holistic framework to solve 
this problem based on an extensive study of processing pipelines. 
Our goal is two-fold: 

\begin{itemize}
  \item Investigate the role of processing pipelines in CSM.
  \item Derive from our results a framework to build holistic training Cover databases.
\end{itemize}

Motivated by these facts, this paper is the first attempt to address the CSM issue by proposing a framework for the generation of databases for holistic steganalysis.
Our work has the following contributions: 

\begin{itemize}
  \item We show that a wise selection of development pipelines allows us to generalize the performances of a steganalysis model on several SOTA databases with fewer training samples.
  \item We also provide a simple yet efficient greedy algorithm for the selection. 
\end{itemize}
 

The rest of the paper is organized as follows:
In Section~\ref{sec:formalization}, we formalize our objective and formulate 
the source selection problem as a set covering problem.
Then, in Section~\ref{sec:experiments}, we present some experiments 
where we want to better understand the origins of CSM meanwhile
finding interesting databases for our battle against CSM. 
Afterwards, in Section~\ref{sec:tackling_csm}, we test the potential
of these bases in a state-of-the-art framework in steganalysis.
Finaly, Section~\ref{sec:conclusion} concludes this work.



%% file: formalization.tex

\subsection{Formalization}

Following \cite{sepak}, we consider that a processing pipeline
is entirely defined by a vector $\omega \in \Omega$ which
contains all the pipeline parameters 
(demosaicking algorithm, 
denoising coefficient, JPEG quality factor, etc.). 
In the steganalysis context, we also introduce a parameter 
$\gamma$ representing steganographer choices,
notably embedding strategy and payload. The state of the art 
for this task essentially lies on machine learning 
models that can be seen as predictors 
\begin{align*}
  f(x \mid \theta_{\omega,\gamma}) : \ & \mathcal{X} \rightarrow \{cover,stego\} \\
  & x \mapsto y
\end{align*}

\vspace*{-0.1cm}
where $\theta_{\omega,\gamma} \in \Theta$ contains all the parameters learnt with covers derived 
from the pipeline $\omega$ and potentially embedded following $\gamma$.

To assess CSM properly, two relevant metrics have been introduced in \cite{giboulot} and \cite{sepak} : 
\begin{itemize}
  \item The Intrinsic Difficulty of a source that is, the probability of error $P_E$ we obtain after
  training on images from this source and evaluating on images from this same source.
 
  \begin{myequation}
    \mathbb E_{(x,y) \sim P((x,y)| \omega,\gamma)} (f(x \mid \theta_{\omega,\gamma} ) \neq y)
  \end{myequation}

  \item The Regret $R_{s,t}$ between two cover sources $s$ and $t$ defined as the difference between the $P_E$ we obtain by training on $s$ and evaluating on $t$ and the Intrinsic Difficulty of $t$. 
  {\large
  \begin{align*}
    \mathbb E_{(x,y) \sim P((x,y)| \omega_t,\gamma)} (f(x \mid \theta_{\omega_s,\gamma} ) \neq y) \\
    - \mathbb E_{(x,y) \sim P((x,y)| \omega_t,\gamma)} (f(x \mid \theta_{\omega_t,\gamma} ) \neq y)  \numberthis
 \end{align*}
  }%
\end{itemize}

Through our study, we are looking for a basis of sources sufficiently rich in order to guarantee a generalization
as great as possible on any source. We formalize this covering objective
as follows : We want a basis $\Omega_B \subset \Omega^B$ s.t.

\begin{myequation}
    \forall \omega \in \Omega,\ \ \exists \omega_b \in \Omega^B \ \ \backslash \ \ R_{\omega_b,\omega} \leq \epsilon \ \
     \label{goal}
\end{myequation}

\noindent
with $\epsilon$ being the maximum level of mismatch we are accepting in terms of Regret. 

\begin{algorithm}[t]
  \footnotesize
  \caption{Greedy covering}\label{alg:cap}
  \begin{algorithmic}
  \Require $\epsilon > 0$, the regret matrix $R$  \\
  Let $N$ be the width of $R$ \\
  Let $Old\_covering$ and $Greedy$ be empty dictionaries
  \For{$i \in N$}
    \State Initialize $P_{\epsilon,i}$ as the set of sources 
    with which we get a regret of at most $\epsilon$ when we train on the $i$-th source 
    \State Fill $Old\_covering[i]$ with $P_{\epsilon,i}$
  \EndFor \\
  \#At that stage we have an initial covering and we are going to refine it.
  \While{$Old\_covering$ is not empty} 
  \State Fill $Greedy$ with the source k that is currently covering a maximum of other sources in $Old\_covering$ : 
   $Greedy[k]  \leftarrow  Old\_covering[k]$
  \For{$i \in N$}
    \State Delete from $Old\_covering[i]$ the sources already covered by $Greedy[k]$ : 
    $Old\_covering[i] \leftarrow Old\_covering[i] \ \backslash \ Greedy[k]$
  \EndFor 
  \EndWhile
  \Ensure $Greedy$
  \end{algorithmic}
\end{algorithm}

\subsection{Extracting reference sources using set-covering}

Given a steganalysis detector and a finite number of sources $N$, (\ref{goal}) can be 
rewritten as the famous set-covering problem \cite{set_covering}.
For each $i \in N$, let $P_{\epsilon,i}$ be the set of all the sources 
with which we get a regret of at most $\epsilon$ when we train on the $i$-th source. 
Starting from the covering $C=\bigcup_{i \in N} P_{\epsilon,i}=N$, we precisely want
to extract a minimal 
subset $N_{\epsilon} \subset N$ such that 
$$C=\bigcup_{i \in N_{\epsilon}} P_{\epsilon,i}=N$$

This problem is NP-complete and therefore, finding the optimal covering $N_{\epsilon}^*$ is not
an easy task. However, a greedy algorithm with a theoretical
upper bound exists \cite{set_covering}. The pseudo-code of this algorithm is presented in Alg.~\ref{alg:cap}. The algorithm first selects the pipeline with the largest cover-set following the constraint on the regret, then append other pipelines not yet covered using a greedy strategy.



%% file: experiments.tex

\subsection{Experimental protocol}

\begin{figure*}[!t]
  \centerline{
  \includegraphics[width=0.95\textwidth]{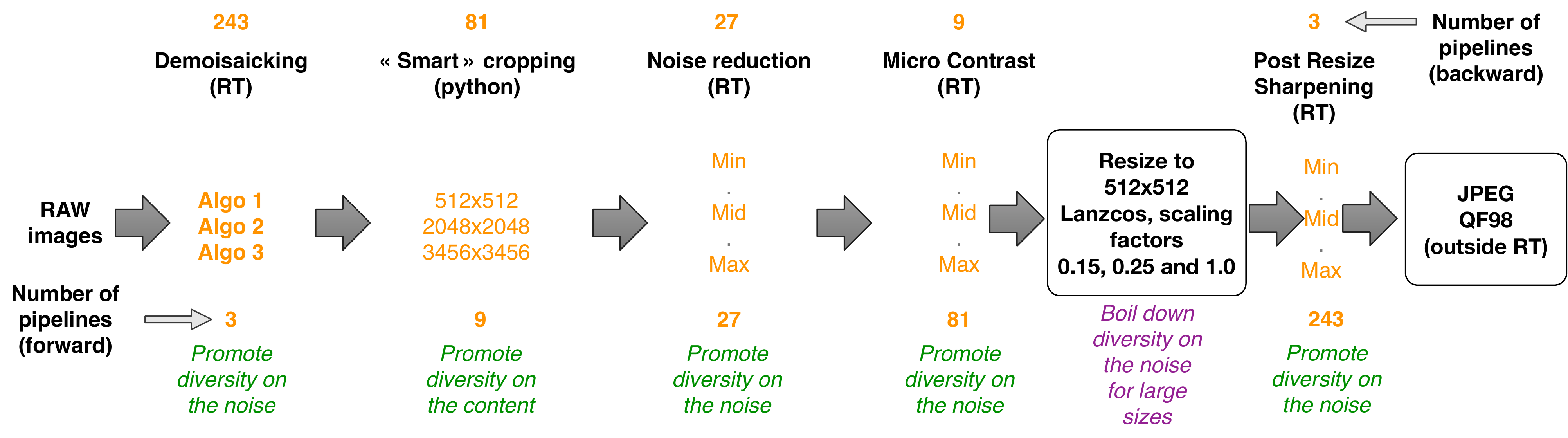}
  }
  \caption{Generation of $3^5$ pipelines. Note that the different scaling factors are only due to crops of different sizes.}
  \label{fig:pipelines}
\end{figure*}

For the experiments presented here, we extract 1115 RAW Images of size $5184\times3456$ from 
ALASKA coming from the camera CANON-EOS-100D and captured with 
ISO $>1000$. The choice of using one given camera model is not innocuous. 
We want to precisely study the role of processing 
pipelines in CSM and hence we are trying to avoid other
important factors such as the sensor quality. Moreover, 
because RAW images associated with high ISO are noisier than low ISO images, this increases the contrast between the different pipelines.

To simulate processing pipelines, we are using 
RawTherapee\footnote{\href{https://www.rawtherapee.com/}{rawtherapee.com}},
an open-source software that handles a 
large range of processing operations ordering from demosaicking to 
jpeg compression. 

The impact of jpeg compressions on steganalysis is already well-known 
in the literature \cite{giboulot}. Here, we propose to put at the end of our processing pipelines a JPEG compression with a 
quality factor of 98 to promote CSM. 

Concerning the head of the pipeline, we 
cherry-picked five operations based on their inherent 
ability to promote diversity in content or noise and, we study $3^5$ combinations 
of them to better understand the mismatch they could bring together. The details about 
covers generation is presented 
in Fig \ref{fig:pipelines}. Please note that we didn't use 
the regular cropping of RawTherapee considering that it leads to low-textured covers
in practice. Instead, the "smart" cropping released on ALASKA2 Challenge
\cite{alaskabase} that is seeking crops with highly textured areas was preferred. It's also important 
to have in mind that this cropping operation may lead to downsampling according to its size. 
Furthermore,
the final JPEG compression is done using Imagemagick\footnote{\href{https://imagemagick.org/}{imagemagick.org}}
in order to fully control it.
The details of the pipelines numbered from 0 to 242 are available 
in our github repo.

From the beginning, the RAW images are randomly split into 50\%
 train and 50\% test.
Afterward, the covers are generated and their embeddings are done
 using UERD \cite{uerd} with a payload 
of 1.5bpp. To keep things simple and to save computational resources, we then
train linear classifiers using DCTR features \cite{dctr} from our sets of covers and stegos.
This payload may seem high but, in practice, it enables obtaining 
cover sources with rather small Intrinsic Difficulty, ranging from 0\% to 14\%. Note that using a standard payload of 0.4bpnzac, it 
resulted in the generation of many cover sources with an important Intrinsic Difficulty, i.e. $\simeq 50\%$. In such a case, the detector is not learning anything and cannot generalize well on other sources, making us blind to the generalization potential of the training source.

Once the cover distributions are generated and the detectors trained on each of them,
we study CSM using a regret matrix $R$ where $R[s,t]=R_{s,t}$ , $s,t \in N^2=\{0,...,242\}^2$. 
Using $R$, we apply then the greedy algorithm presented in Alg.~\ref{alg:cap} with 
$\epsilon=2\%, 4\%,..., 10\%$. In Table~\ref{tab:covering}, you can find 
some information about the covering obtained and, in Table
\ref{tab:regret_10} the regret matrix of the 5 sources returned for $\epsilon=10\%$.

\begin{table}[h]
    \centering
    {
    \begin{tabular}{|c|c|c|}
    \hline
    $\epsilon$ & $\vert N_{\epsilon}\vert$ & $\min$  \\
     &  & $\vert N_{\epsilon}^*\vert$ \\ \hline
    2\%        & 26            & 6        \\ \hline
    4\%        & 14            & 3        \\ \hline
    6\%        & 11            & 3        \\ \hline
    8\%        & 10            & 2        \\ \hline
    10\%       & 5             & 1        \\ \hline
    
    \end{tabular}
    }%
    \vspace*{0.2cm}
    \caption{Covering size obtained using Alg.~\ref{alg:cap} for different
    values of $\epsilon$. The right column is the minimum possible size of the covering-set (see~\cite{set_covering}).}
    \label{tab:covering}
\end{table}

\vspace*{-0.2cm}

As one can expect, the lower $\epsilon$, the more sources 
we need to guarantee a regret less than $\epsilon$ for everyone. Moreover,
we also observe an interesting property for the sources returned 
by the greedy algorithm. By inspecting Tab~\ref{tab:regret_10}, the extracted representatives are "complementary" since they are associated with important regrets regarding all the other selected sources. This feature is expected since the selected sources are representing sources of different types.

\begin{table}[!h]
  \centering
  \begin{tabular}{|c|c|c|c|c|c|}
  \hline
  Train / Eval & \textbf{21} & \textbf{22} & \textbf{31} & \textbf{60} & \textbf{229} \\ \hline
  \textbf{21}  & 0           & 20          & 12          & 30          & 30           \\ \hline
  \textbf{22}  & 5  & 0           & 6           & 25          & 37           \\ \hline
  \textbf{31}  & 21 & 10          & 0           & 32          & 33           \\ \hline
  \textbf{60}  & 25 & 26          & 20          & 0           & 30           \\ \hline
  \textbf{229} & 19          & 16          & 29          & 32          & 0            \\ \hline
  \end{tabular}
  \vspace*{0.3cm}
  \caption{Regret matrix (in \% ) between 
    the 5 sources in $N_{10\%}$}
  \label{tab:regret_10}
  \end{table}

\subsection{Analysis: which parameter promotes heterogeneity?}

\begin{figure}[!h]
  \centerline{
  \includegraphics[width=\columnwidth]{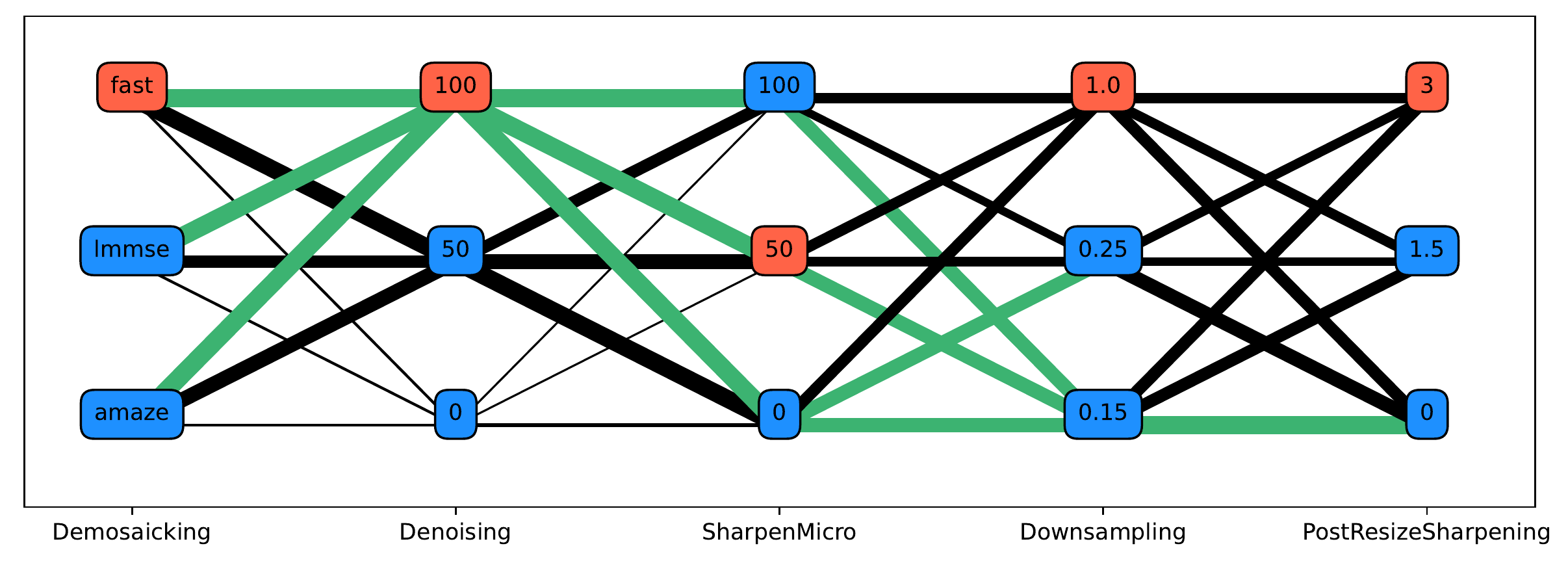}
    }
    \vspace*{-0.2cm}
  \caption{Content of the sources covered by pipeline \#229 in $N_{10\%}$. 
  The red boxes are the parameters of \#229 \& $\vert C_{\#229} \vert = 159$ sources.
  The green links are the most represented links among the 9 possible at each stage.}
  \label{fig:cluster_10_229}
\end{figure}

\begin{figure}[!h]
  \centerline{
  \includegraphics[width=\columnwidth]{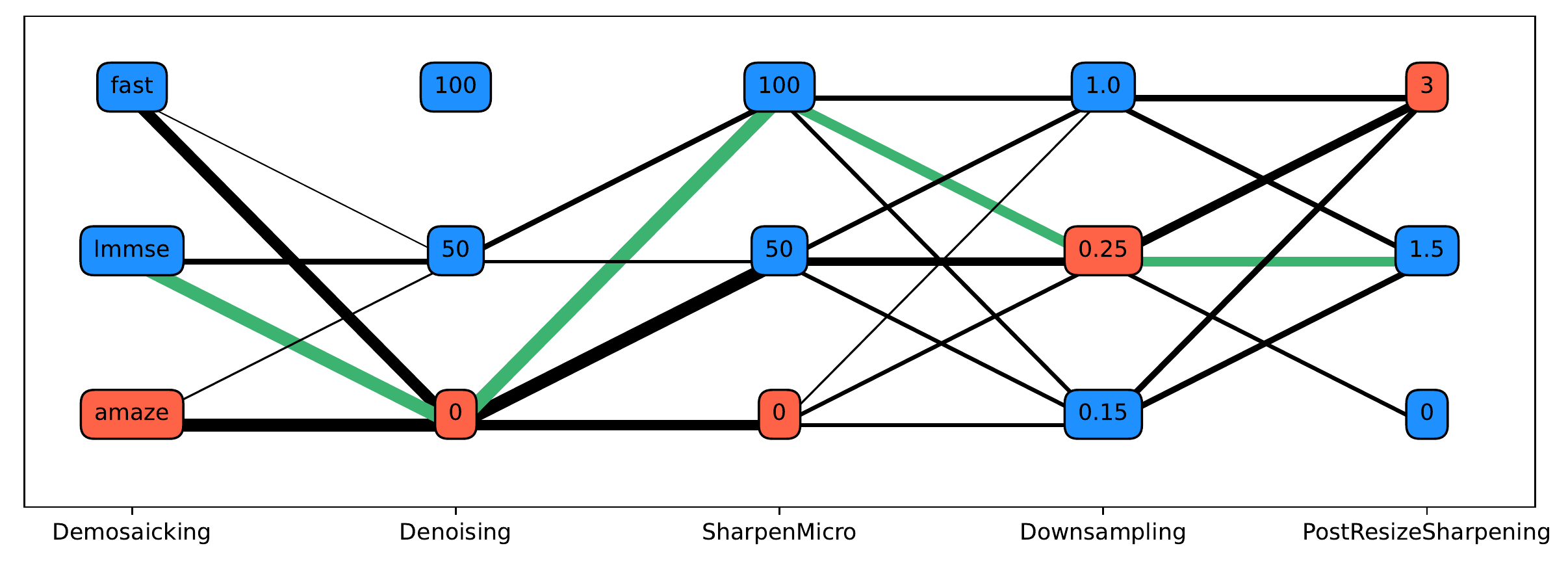}
    }
    \vspace*{-0.2cm}
  \caption{Content of the sources covered by pipeline \#60 in $N_{10\%}$.The red boxes are the parameters of \#60 \& $\vert C_{\#60} \vert = 69$ sources.
  The green links are the most represented links among the 9 possible at each stage.} 
  \label{fig:cluster_10_60}
\end{figure}

Once the greedy algorithm returns a source covering, we assign 
labels to each of our 243 sources according to the representatives in $N_{\epsilon}$ that
cover them with a radius $\epsilon$. In doing so, we are 
disclosing clusters of sources that help us to realize which parameters of 
our processing pipelines are the most discriminating. For instance, for $N_{10\%}$ we end up with 
5 clusters and the covering follows a kind of Pareto law: Two clusters are encompassing 94\% of the sources. In Figures \ref{fig:cluster_10_229}
and \ref{fig:cluster_10_60} we present visually their substance.

From Fig.~\ref{fig:cluster_10_229} we realize that pipeline \#229 is issued from important denoising followed by 
an important adding of noise. Concretely, the sensor noise has been well cleaned and an artificial noise is 
then added based on what is remaining. Surprisingly this intriguing combination enables to cover around 
66\% of our arsenal of 243 pipelines with a 10\% regret radius. The pipelines covered are mostly the ones issued from 
high denoising. Pipeline \#60
is issued from a high cropping 
factor, hence followed by important downsampling and, a significant adding of noise. This time the sensor 
noise is not cleaned, we are extracting more content and we end the processing by adding artificial noise through sharpening. This mix enables us to cover around 28\% of our sources, corresponding roughly to most noisy 
sources. 

Without surprise, decreasing the maximal level of regret wanted allows one to blatantly reveal the most difficult sources to cover 
very precisely. The more we decrease $\epsilon$, the more we are building sparse clusters, and the more 
we can observe what parameters make very singular and specific sources.

\begin{figure}[!h]  
  \centerline{
  \includegraphics[width=\columnwidth]{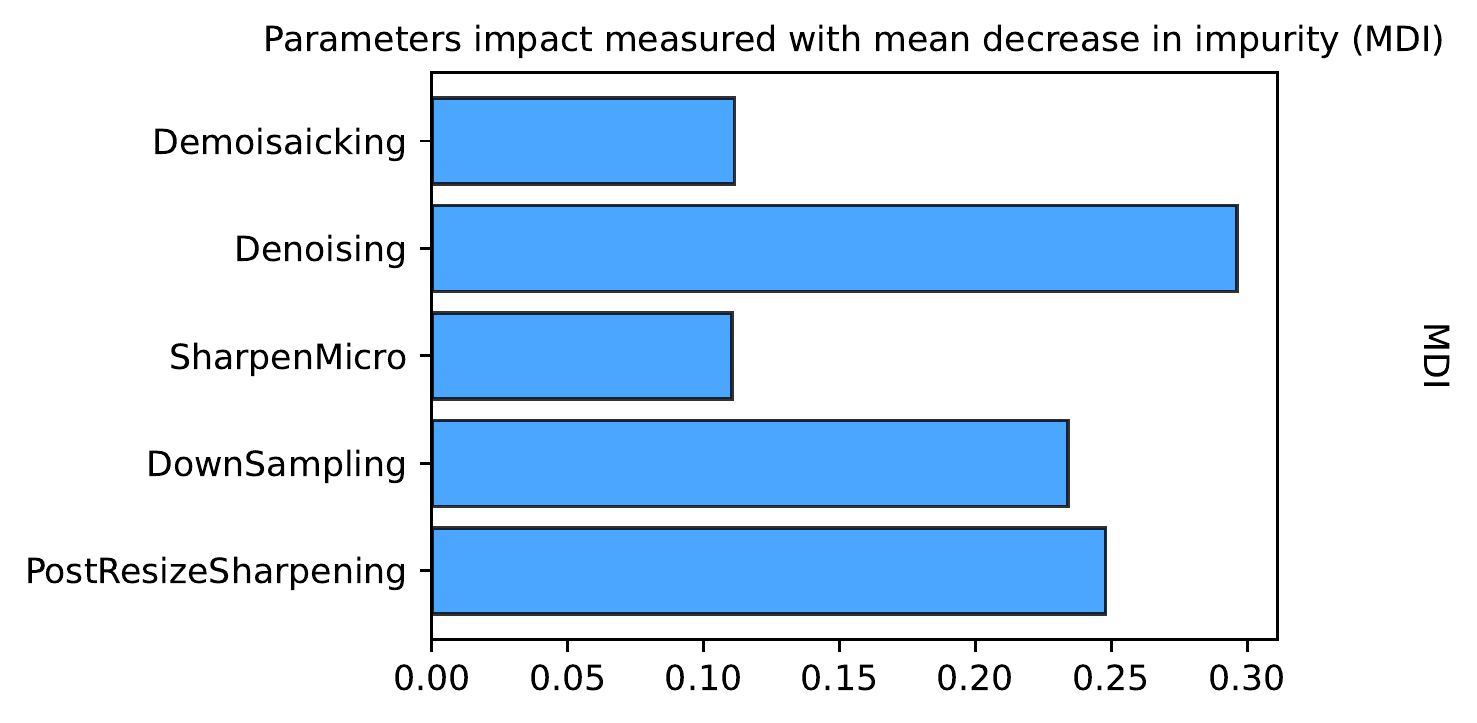}
    }
    \vspace*{-0.5cm}
  \caption{Study of parameters importance in the creation of singular sources \\ \centering (using $N_{1\%}$)}
  \label{fig:MDI}
\end{figure}

Using $\epsilon=1\%$, we have 30 clusters, most of them with very few members. 
Instead of analyzing each cluster, we decide this time to train a random forest algorithm with an entropy
criterion, trained to guess the clusters of each source according to their pipeline parameters. In Fig. \ref{fig:MDI},
we present the importance of each parameter in the decisions of the model thanks to the Mean Decrease Impurity (MDI), which expresses the classification gain associated with splits of the forest for a given variable~\cite{breiman2002manual}.

Fig \ref{fig:MDI} shows that Denoising, PostResizeSharpening, and Downsampling are the parameters that 
play the biggest roles in the generation of singular sources. The Demosaicking and 
the SharpenMicro operations seem much less significant. These observations are consistent with the work 
of \cite{giboulot} and Figs \ref{fig:cluster_10_229} \& \ref{fig:cluster_10_60}.


%% file: tackling_csm.tex
\subsection{Reference framework}

At this point, we extracted some sets of sources enabling us to generalize on the grid, up to 
predefined levels of regret.  However, this ability to generalize may also depend on the embedding strategy 
and the detector used. Hence, we propose now to test the potential of our set of sources in a reference 
framework representing the current state-of-the-art.

The pre-trained J-UNIWARD ImageNet (JIN) \cite{pretrainforstega} achieves currently state-of-the-art results 
on classical steganalysis databases for the J-UNIWARD \cite{juniward} embedding with a payload between 0.4 and 0.6
bpnzac. Hence, we propose to embed our covers using J-UNIWARD with a payload of 0.5bpnzac and then fine-tune JIN 
for our experiments. To save computational resources while achieving a satisfying convergence, we fix a maximum cover training budget of 80K $256$x$256$ images. The training, validation, and testing sets are split as follows 80/10/10. All the  
hyperparameters regarding the optimizer and the scheduler are fixed following \cite{pretrainforstega}.
This setup is the one adopted for all the upcoming experiments presented in this paper.

As mentioned previously, we systematically 
observed a kind of Pareto law for our different coverings. For $\epsilon \leq 10\%$, there are singular sources,  
often presenting a high Intrinsic Difficulty, that are covering only a few others. During our experiments,
we noticed that these specific sources are impacting negatively training with JIN. Discarding representatives that 
are not covering at least 10 sources revealed to be an effective strategy 
enabling to disclose the potential of the most interesting ones. We call RT$_\epsilon$ the bases resulting from 
this filtering strategy.

We also create bases made of pipelines picked randomly so that we can compare the results 
obtained with our greedy coverings and randomness. For each $\epsilon$ selected, we pick randomly as many pipelines
like the ones in RT$_\epsilon$ and, we call RANDOM$_{\epsilon}$ the source resulting from the mix of these pipelines.
For each $\epsilon$ we create 3 variants of RANDOM$_{\epsilon}$ to be able to compute 
a mean performance of the random strategy.

\begin{table*}[t]
    \centering
    
    \resizebox{\linewidth}{!}{%
    \begin{tabular}{|c|c|c|c|c|c|c|c|c|c|c|} 
    \hline
    \textbf{\textbf{Eval \textbar{}}\textbf{Train}} & \thead{IMAGENET \\ (860K)}& \thead{RTBASE \\ (64K)} & \thead{ALASKA \\ (64K)} & \thead{BOSSBOWS \\ (64K)} & \thead{FLICKR \\ (64K)} & \thead{RT$_{2\%}$ \\ (18K)} & \thead{RT$_{4\%}$ \\ (18K)} & \thead{RT$_{6\%}$ \\ (14K)} & \thead{RT$_{8\%}$ \\ (11K)} & \thead{RT$_{10\%}$ \\ (7K)} \\ 
    \hline
    \thead{RTBASE (16K)} & 11 & 0 & 1 & 5 & 8 & 1 & \textbf{0} & 1 &  3 & 3 \\ 
    \hline
    \thead{ALASKA (16K)} & 9 & 3 & 0 & 4 & 7 & \textbf{2} & \textbf{2} & \textbf{2} & 5 & 5 \\ 
    \hline
    \thead{BOSSBOWS (16K)} & 10 & 6 & \textbf{5} & 0 & 13 & 8 & 9 & 8 & 13 & 10 \\ 
    \hline
    \thead{FLICKR (16K)} & 20 & 18 & 19 & 16 & 0 & 18 & 15 & 18 & \textbf{10} & 13 \\
    \hline
    \end{tabular}
    }    
    \centering
    \caption{Regret matrix : Results on our bases of interest (in \%).}
    \label{tab:global_results}
    \end{table*}
    
    \begin{table*}[t]
        \resizebox{\linewidth}{!}{%
    \begin{tabular}{|c||c|c|c||c|c||c|c||c|c|} 
        \hline
        \textbf{\textbf{Eval \textbar{}}\textbf{Train}} & RT$_{2\%}$ &  RT$_{4\%}$ & Random$_{2\% \& 4\%}$ & RT$_{6\%}$ & Random$_{6\%}$ & RT$_{8\%}$ & Random$_{8\%}$ & RT$_{10\%}$ & Random$_{10\%}$   \\ 
        \hline
        RTBASE    & 1 & \textbf{0} & 2 &   \textbf{1} &  2 &  \textbf{3} & \textbf{3} &  \textbf{3} & 10\\ 
        \hline
        ALASKA & \textbf{2} & \textbf{2} & 3 &   \textbf{2} &  4 &  \textbf{5} & \textbf{5} &  \textbf{5} & 16\\ 
        \hline
        BOSSBOWS  & 8 & 9 & \textbf{6} &   8 &  \textbf{7} &  13 & \textbf{7} &  10 & \textbf{9} \\ 
        \hline
        FLICKR   & 18 & \textbf{15} & 18 & \textbf{18} & \textbf{18} & \textbf{10} & 17 & \textbf{13} & 19 \\
        \hline
        \end{tabular}
    }
    \vspace*{0.3cm}
    \caption{Regret matrix : Comparison between covering and random selection of pipelines (in \%).}
    \label{comp_with_random}
    \end{table*}

\begin{table}[b]
    \centering
    \resizebox{\linewidth}{!}{%
    \begin{tabular}{|l|l|l|l|} 
    \hline
     RTBASE & ALASKA20K & BOSSBOWS & FLICKR \\ 
    \hline
     27\% & 28\% & 13\% & 24\% \\
    \hline
    \end{tabular}
    }
\vspace*{0.3cm}
\caption{Intrinsic Difficulties ($P_E$) of our evaluation bases.}
\label{Intrinsic_difficulty}
\end{table}

\subsection{Bases of interest}

Fine-tuning JIN using the main sources in $N_\epsilon$ for some $\epsilon$, we would like to see how much we can 
generalize on different bases of interest. Since we didn't include the JPEG
compression in our previous analysis, we decide to only work with pictures compressed at the same quality factor as ours,
namely, 98. We propose for instance, the following bases:
\begin{itemize}
    \item A base of 80K covers generated with all our $3^5$ pipelines (RTBASE). 
    \item The 80K covers resulting from the concatenation of BOSS \cite{bossbase} \& BOWS-2 \cite{bows} compressed with a QF of 98 (BOSSBOWS).
    \item 80K of covers chosen randomly from ALASKABASE \cite{alaskabase} compressed with a QF of 98 (ALASKA). 
    \item 80K of "wild covers" compressed with a QF of 98 derived from unknown realistic pipelines.
\end{itemize}

For our wild base, we propose to build it using the well-known
website flickr\footnote{\href{www.flickr.com}{flickr.com}} which gathers 
images shared by millions of users. Considering that we don't have any idea about the processing pipelines used 
by the users, we are in the worst-case scenario. We are notably completely blind about the quantization tables used
for the compressions and our JPEG-based filtering is hence, approximative. 
Moreover, to focus on CSM mostly caused by unknown processing pipelines, we prefer to select images coming from camera models in the same  
range as the one we used for the grid (CANONEOS).

We share in \ref{Intrinsic_difficulty} the Intrinsic Difficulties of all the bases described above, finetuning JIN
over 15 epochs. All our evaluation sources are reasonably difficult. This is possible thanks to the quality of 
the pre-trained weights released by \cite{pretrainforstega}. We tried at first to perform trainings from scratch without 
these weights. Unfortunately, it wasn't possible to make JIN converge properly with a classic training, that's to say,
a training that does not involve any steganalysis tricks like the pair constraint \cite{pretrainforstega}.

\subsection{Results}

In Table \ref{tab:global_results}, we present 
the regret matrix on the bases of interest obtained after fine-tuning JIN over 15 epochs. The first 
column is a special case where we present the regrets without fine-tuning, simply using the pre-trained weights computed using ImageNet \cite{ImageNet,pretrainforstega}. As a first result,
we can already observe that using JIN "on the shelf" is not the best strategy if we want to generalize on images not coming from ImageNet. That being said, as explained before, we cannot neglect the positive effect of these pre-trained weights on our trainings.

The BOSSBOWS test database is more challenging for the detectors trained on our extracted bases RT$_{\epsilon}$. This is partly because we started our study with a set of images captured with High ISO, a pattern not very represented in BOSSBOWS which gathers images taken with rather low ISOs. 

It is also interesting to notice that, despite 
the drastic change in framework, 
the bases RT$_{\epsilon}$ still enable ensure a regret lower than $\epsilon$ on the grid represented by RTBASE. Moreover, for $\epsilon = 2,4,6\%$, we obtain a pretty low regret with ALASKA. This 
is rather surprising considering the great diversity in terms of sensor and ISO of the images contained in ALASKA that is far more 
superior than our bases. This indicates again that learning with a wide noise diversity is more important than learning with a wide content diversity if we want to be as holistic as possible. Moreover, we obtain comparable performances on ALASKA with respect to RTBASE, even if, fewer training samples and fewer pipelines are used to train the RT$_{\epsilon}$-bases compared to RTBASE.

Looking at the performances on FLICKR, three of our five 
extracted bases are outperforming all the other training bases, even in the case where they are made of more samples. The case of RT$_{8\%}$ is particularly
interesting since it is a result we cannot reproduce using other source combinations.
This shows on its own the importance 
of cherry-picking the pipelines used for the training base instead of using as many images as possible, or making a random mixture with a maximum of pipelines.

At last, we present in Tab. \ref{comp_with_random} a performance comparison  after training on RT$_{\epsilon}$ and Random$_{\epsilon}$. Except for BOSSBOWS,
it seems that using our bases is a better strategy than generating random combinations of pipelines, especially if we want to generalize on FLICKR.

%% file: conclusion.tex
In this paper, we present a methodology for generating relevant bases 
enabling to fight CSM in an operational steganalysis framework. 
We show that our strategy leads to bases more informative for a 
detector than exploiting a high quantity of images, random augmentations, or 
as many pipelines as possible. Furthermore, from our different studies, it appears that Denoising, Sharpening, and Downsampling are playing a significant role in the cover source mismatch issue. Broadly speaking, other studies should be conducted to fully harness these parameters to cope with CSM. 
From our study, it is also easy to derive a batch of "complementary" sources. This may help the 
community to test strategies that are trying to reduce the mismatch between a 
set of sources. In the near future, we plan to perform such experiments 
leveraging domain adaptation methods just like we already did in \cite{abecidan} for 
digital forensics.